# A Practical Methodology for ML-Based EM Side Channel Disassemblers

Cesar Arguello
CISE
University of Florida
carguello1@ufl.edu

Hunter Searle
CISE
University of Florida
huntersearle@ufl.edu

Sara Rampazzi
CISE
University of Florida
srampazzi@ufl.edu

Kevin Butler
CISE
University of Florida
butler@ufl.edu

*Abstract*—Providing security guarantees for embedded devices with limited interface capabilities is an increasingly crucial task. Although these devices don't have traditional interfaces, they still generate unintentional electromagnetic signals that correlate with the instructions being executed. By collecting these traces using our methodology and leveraging a random forest algorithm to develop a machine learning model, we built an EM side channel based instruction level disassembler. The disassembler was tested on an Arduino UNO board, yielding an accuracy of 88.69% instruction recognition for traces from twelve instructions captured at a single location in the device; this is an improvement compared to the 75.6% (for twenty instructions) reported in previous similar work.

## 1. Introduction

Embedded devices form an integral part of the modern computing ecosystem. They can be found in a myriad of applications, ranging from household appliances to security-critical industrial controllers. Securing this wide range of devices is a massive and crucial design challenge, especially with the rise in connectivity of the Internet of Things and emerging threats [1]. Many of these devices, such as fuel tank monitors or farming field sensors, offer little insight into their internal workings due to proprietary technology or limited interfaces. Additionally, devices such as medical devices might be deployed long-term, with little or no ability to update their software against cyberthreats. This combination of longevity and minimal access create a situation where devices are susceptible to many forms of attacks, including disrupting functionality, falsifying sensor output, and increasing power consumption to drain batteries [2].

One proposed approach to the challenge of protecting embedded devices against these attacks is to use side-channel information. Side-channels refer to information that is leaked by unintentional signals generated in the normal operation of a processor. There are many forms of side-channel, including power, noise, electromagnetism, timing, etc [3] [4] [5]. In each case, the signal is correlated to the operations that generated them, so that information about those operations can be retrieved from the signal. Of these, power side-channels offer the clearest signal, and have consequently received more attention from researchers. However, they require a direct connection to the processor's power rail. Electromagnetic signals, however, can be captured with no physical connection, making them more preferable as a side-channel security mechanism.

In this work we propose an efficient approach for building an electromagnetic side-channel based disassembler, which can identify specific instructions being executed on an embedded processor. We then use our approach to build a proof-of-concept EM-based disassembler to be used for implementing anomaly detection and control flow mechanisms on low-cost embedded systems without the need for firmware or hardware redesign. In comparison with previous work [6], [7] that uses wiring or multiple measurement points, our approach leverages the presence in the device's board of electronic components that generate EM emanations (e.g., amplifiers) correlated to the internal processor operations. Using a simple random forest algorithm for classification, we achieve single instruction granularity with 88.69% accuracy over twelve different instructions on an ATMega328P 16MHz processor. This is an improvement over previous work [6] which required multiple measurement points and was only tested on a 4MHz processor to achieve 75.6% accuracy on 20 instructions. These encouraging results open the possibility of utilizing our approach to build more efficient and high-accuracy disassemblers to be used for anomaly detection.

## 2. Background

Side-channel signals have been a known source of information leakage for decades [8] [9] [10]. They have been exploited for recovering different types of data, including screen images [10], secret keys [3], and audio [11]. Only recently have side-channel analysis techniques been turned to program disassembly, starting when Eisenbarth et al. built a disassembler using power analysis [12]. Since then, much progress has been made in this research area. SCANDALee was the first successful em-based disassembler [6]. Their method required multiple measurement locations, and could only recover a portion of the instruction set on a slower (4MHz) processor. New techniques were introduced for disassembly by Park et al., who were able to recover nearly the entire instruction set, along with registers, of a faster processor (ATMega 238P) [7], although they only consider power side-channels, which require a wired connection. Neural networks have also been leveraged for disassembly, achieving similar levels of accuracy only on power side channel [13]. Other works have sought to ensure security without instruction-level disassembly, such as Khan et al.'s EM-based intrusion detection system [14]. Our methodology builds on these previous works to overcome the limitations of multiple-point measurement and signal extraction to achieve instruction-level disassembly using EM emissions.

## 3. Methodology Overview

Our process is composed of three steps: 1) Leakage detection, 2) Signal extraction, and 3) Classification. Leakage detection is only done once, whereas signal extraction and classification are performed every time.

**Leakage Identification.** Following the template-based method used by previous work [7], we sample individual instructions to build a template model before classifying them in real code. Our approach begins by identifying the component from which EM emanations are most strongly correlated to executing instructions. This is done empirically with a grid search across the device PCB using a EMI probe located a few centimeters from the board. At each point in the grid, we perform a simple classification using a Random Forest algorithm that has been adapted for use with time series implemented by sktime [15]. To use this method we generate a template built from samples of individual instructions with a no-operation (NOP) immediately preceding and following it to better identify the signal correlation with the executed instructions. Before and after the target instruction, we also implement a trigger operation for inducing a voltage change (e.g., an instruction to flip a GPIO bit). This trigger is used to separate the individual instructions for classification. This procedure gives a lower single-instructions accuracy than our final classification, however, it shows the relative information leakage that is used to identify one or more optimal measurement locations corresponding to the "leaky" electronic components (see Figure 2). After this procedure the optimal measuring point is selected and used for the rest of the process. Finally, because our process uses the Fast Fourier Transform of the signal, rather than the time series data, we identify the target frequency bands using a spectrum analyzer.

**Signal Extraction and Classification.** The process to build our template and to classify unknown signals follow the same steps of signal extraction and classification. This step consists in measuring the magnetic field around the optimal point identified in the first step. Then we automatically separate the instructions using the triggers, and compute the FFT for each instruction. The magnitudes of the transform at the frequencies identified in step one are then used as features to train our model. While our methodology can be adapted to any machine learning algorithm, in our case study we found that the Random Forest algorithm (not to be confused with the Time Series Random Forest algorithm that we used for the grid search) gave the highest single-instruction accuracy.

## 4. Proof-of-concept Disassembler

To test our methodology, we built a proof-of-concept disassembler. We selected a subset of the AVR instruction set by examining real-world code for a stack overflow attack, then tested our disassembler on an ATMega328P.

**Experimental setup.** Our acquisition setup is shown in Figure 1. We used a Tektronix H10 H-Field Probe connected to a Tektronix MDO4024C oscilloscope to collect traces from an Arduino UNO.

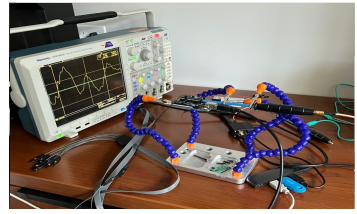

Figure 1. Acquisition setup made of a Tektronix H10 H-Field Probe connected to a Tektronix MDO4024C oscilloscope.

**Leakage Identification Phase.** We began with a 80-point grid search over the Arduino board and ran the Time Series Random Forest classifier. We found that the highest accuracy was near an operational amplifier connected to the crystal oscillator, shown in Figure 2. This was the point at which all other measurements were taken.

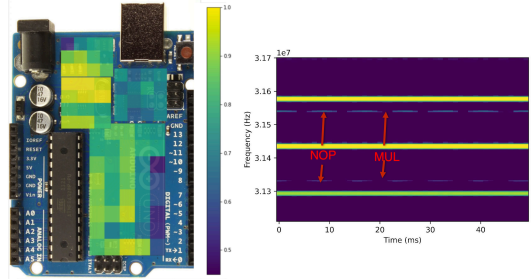

Figure 2. (Left) Recognition accuracy for the 80 subsection in which the device under test was divided. The area near to the operational amplifier connected to the crystal oscillator shows the highest accuracy (yellow). (Right) Example of spectral profiling on a group of MULs and NOPs.

We used the Signal Hound SA44B spectrum analyzer to study the frequency spectrum. Because the leakage component was closer to the Arduino 16 MHz clock, we examined frequencies within 1 MHz of the clock sub-harmonics. Our analysis found that frequencies between 31.3-31.6 MHz showed the largest voltage difference for our test instructions. The bands are shown in Figure 2. Finally, we performed the hyperparameter optimization for the tested classification algorithms as shown in Table 1. The random forest algorithm presented the highest accuracy and was hence chosen as the optimal one for the development of the disassembler.

TABLE 1. RANDOM SEARCH HYPERPARAMETER OPTIMIZATION

| Algorithm | Accuracy | Optimum Hyperparameters |
|---|---|---|
| Random Forest | 85% | Num. Estimators = 1000  Min. Interval = 2 |
| Random Interval Spectra | 56% | Num. Estimators = 829  ACF lang = 400 |
| K Nearest Neighbors | 76% | Num. Neighbors = 100 |
| Support Vector Machines | 82% | Kernel = linear  Gamma = 0.1 |

## 4.1. Evaluation

**4.1.1. Single Instructions Recognition.** The results of accuracy recognition for twelve selected instructions after performing four fold cross validation in the developed disassembler are summarized in Figure 3. The proposed implementation yielded an 88.69% recognition accuracy.

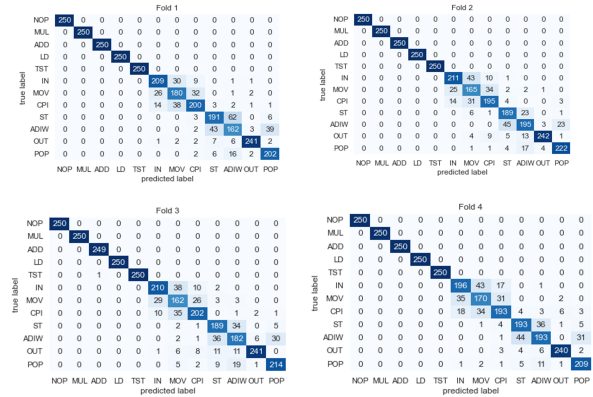

Figure 3. Confusion matrix of our EM side-channel based disassembler four cross validated

**4.1.2. Code Recognition.** To test the capability of our approach to recognize instructions on a potential real-world case study, we consider the scenario of a real-world medical device, SyringePump [16]. The pump is designed to deliver medications to a patient at periodic intervals. The system typically consists of a syringe, an actuator (a stepper motor), and a control unit (Arduino UNO) that takes commands from the serial port. We focus on a specific section of the Arduino SerialRead function as test case. This function is crucial for anomaly detectors to infer the presence of ongoing buffer overflow attacks on the internal buffer of the serial port.

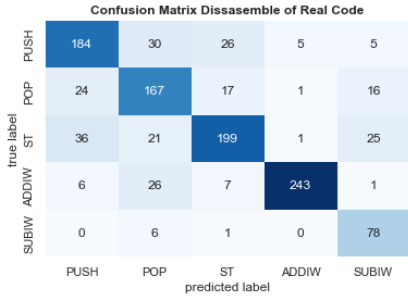

Figure 4. Confusion matrix for identification of instructions in three cycle instructions of the partially implemented Arduino SerialRead function

A small driving program was made such that a computer sends three arbitrary characters to the Arduino board serial port and the SerialRead function stores those characters in a buffer. Our acquisition setup was used to collect traces for the relevant five instructions used by the function. The driving program was run 500 times and 75% of these traces were used to improve the single-instruction model while the rest was used for testing. The recognition accuracy obtained was 77.4% as shown in Figure 4.

## 5. Observations and Future Work

The proposed methodology allows single-instruction classification with high accuracy. Moreover, by observing the number of certain operations execution as well as the timing deviation from the regular behavior, an anomaly detector can leverage this findings to automatically determined if an attack was causing such deviation. Although, the results of this test case scenario are preliminary and more testing is required to evaluate the robustness of our approach on different scenarios, they also show the potential of our methodology to be used not only identify anomalies but also provide forensic evidence for their categorization. Planned future work will address a more in-depth evaluation on multiple case scenarios and different processors to fully validate our methodology.

## Acknowledgment

This work was funded in part by The Center for Enabling Cyber Defense in Analog and Mixed Signal Domain (CYAN - AFRL FA8650-19-1-1741) and a gift from Facebook.